\documentclass[conference]{IEEEtran}
\usepackage{graphicx} 
\usepackage{setspace} 
\usepackage{multicol}
\usepackage{multirow}
\usepackage{tikz}
\usepackage{comment}
\usepackage{framed}
\usepackage[T1]{fontenc}
\usepackage[utf8]{inputenc} 
\usepackage{url}

\usepackage{textcomp}    

\usepackage{color}
\usepackage{cite}
\usepackage{algorithm,algpseudocode}

\usepackage{listings}
\usepackage{xcolor}
\lstdefinestyle{sharpc}{language=[Sharp]C, frame=lrtb, rulecolor=\color{blue!80!black}, 
    basicstyle=\scriptsize,
    numbers=left,
    stepnumber=1,
    showstringspaces=false,
    tabsize=1,
    breaklines=true,
    breakatwhitespace=false,}

\usepackage{epstopdf}
\usepackage{fancyvrb}   

\usepackage{tabularx}  

\ifCLASSINFOpdf
\else
\fi
\begin{document}
%

\title{IntegrationDistiller: Automating Integration Analysis and Testing of Object-Oriented Applications}


%

\author{\IEEEauthorblockN{Mehrdad Saadatmand\IEEEauthorrefmark{1}}
\IEEEauthorblockA{
\IEEEauthorrefmark{1}Research Institutes of Sweden (RISE)\\ RISE SICS V\"{a}ster{\aa}s\\}


\author{\IEEEauthorblockN{Mehrdad Saadatmand}
\IEEEauthorblockA{
Research Institutes of Sweden (RISE)\\ RISE SICS V\"{a}ster{\aa}s \\}

Email: mehrdad.saadatmand@ri.se}}



%


\maketitle

\begin{abstract}
Software systems typically consist of various interacting components and units. While these components can be tested and shown to work correctly in isolation, when integrated and start interacting with each other, they may fail to produce the desired behaviors and results. Integration testing plays an important role in revealing issues in interactions among cooperating components. Identifying different interaction scenarios, however, is not a trivial task when performing integration testing. On the other hand, most of the integration testing solutions proposed in the literature are manual which hinders their scalability and applicability when it comes to large industrial systems. In this paper we introduce IntegrationDistiller as an automated solution and tool to identify integration scenarios and generate test cases (in the form of method call sequences) for .NET applications. It works by analyzing the code and automatically identifying class couplings, interacting methods, as well as invocation points. Moreover, the tool also helps and supports testers in identifying timing issues at integration level by automatic code instrumentation at invocation points. The code analysis engine of IntegrationDistiller is built and automated using .NET compiler platform, known as Roslyn. Hence, this work is the first in utilizing Roslyn features for automatic integration analysis and integration test case generation. This work has been done as part of our collaboration with ABB Industrial Automation Control Technologies (IACT) in Västerås-Sweden to address the integration testing challenges of the software part of the ABB Ability\texttrademark~800xA distributed control systems. 

\end{abstract}


%
\IEEEpeerreviewmaketitle

\section{Introduction}\label{sec:into}

With the transition of traditional industrial systems towards more software-based solutions, the role of software in industry has become more and more dominant and important than ever before. Such transitions are already observed in domains such as automotive, telecommunication, and process industry as more functionality and features are implemented and assigned to software components in a product. On the other hand, this transition necessitates more rigorous verification and validation as the quality of the embedded software can now have direct and bigger impact on the quality of the end-product that is delivered to the customers (e.g., a car). This also brings along the following requirements on the testing techniques for ensuring the quality of software products in industry: 
\begin{itemize}
\item scalability: the testing techniques adopted should be scalable with respect to the increasing size and complexity of real-world industrial software applications (e.g., \textit{amount} of code committed, and changes made daily by different development teams located in different parts of the world; as in distributed development environments),
\item automation: as the size and complexity of industrial software products grow rapidly, automation in testing gains an important role particularly related to the scalability issue as manual solutions can be too costly and time-consuming to scale well and be sustainable in the long run. This is particularly important considering the \textit{frequency} of code commits and daily changes made by different distributed development teams as in continuous integration and deployment environments, and when continuous decisions on test prioritization, selection and execution for regression testing need to be made repeatedly,
\item seamless integration: testing techniques should not disrupt and negatively impact the current development process of a company, and should fit seamlessly with it. There are many proposed testing techniques in the literature which may perform better than current ones already used in a company, but are not adopted simply due to this issue.

\end{itemize}

\noindent Also, scalability is considered as one of the main challenges in testing of software systems, and effective automation is considered a prerequisite for scalability\footnote{Lecture on 'Artificial Intelligence for Automated Software Testing', Lionel Briand, ISSTA, Amsterdam, Jul 2018, \url{https://www.slideshare.net/briand_lionel/artificial-intelligence-for-automated-software-testing-106757936}}.

Software systems are typically built as a set of different components and units that interact and work together to provide certain functionality. As the system and its components get \textit{integrated}, such interactions need to be tested. This is particularly important noting that components can be tested separately and in isolation, and come out as correct. However, when they are integrated and expected to interact and work with other components, they may be in conflict with each other and fail to provide the expected behaviors and results. One major challenge in this regard is to determine different combinations and interaction scenarios of system components to be able to test them. More importantly, as discussed above, this needs to be done in an automated way to be scalable in industrial settings. Also generally the larger a project \cite{IntegrationMutation} and the higher the number of distributed teams involved in its development, the more important integration testing in that project will be.

In this paper we introduce IntegrationDistiller as a tool and solution set for automated analysis of object-oriented applications with respect to integration scenarios and automatic generation of integration test cases examining the interactions between classes. At its core and in simple terms, it works by identifying different method call sequences that can result in different states of class objects by analyzing how and where class fields are used and modified. In short, this paper and the tool that we introduce contribute with the following points:

\begin{enumerate}
\item an automated solution to analyze classes, their dependencies and interactions, and generate test cases exercising those interactions,
\item automated analysis of object-oriented source code and determining invocation points,
\item automatic code instrumentation at invocation points to estimate timing properties and enable testing for timing issues at integration level.
\end{enumerate}  

The main novelty of the work thus lies in providing a fully automated approach for integration analysis and generation of integration level test cases based on automatic data-flow analysis. The scalability aspect is addressed in this work through effective automation, as one of the main prerequisites of scalability. The solutions have been developed as part of the research collaboration with ABB Industrial Automation Control Technologies (IACT) in Västerås-Sweden to address the integration challenges of the software part of the ABB Ability\texttrademark~800xA distributed control systems which are described in more detail later in the paper. The rest of the paper is structured as follows. In Section \ref{sec:background}, background context in which the work has been performed along with preliminary introduction to the core concepts used throughout the paper are provided. Section \ref{sec:approach} describes the details of the approach and tool. Since we can not include the actual code from our industrial partner, a demonstrative example using dummy code is provided in Section \ref{sec:example} to show how the approach works and what information and outputs it generates. In Section \ref{sec:discussion} different aspects of the proposed solution are discussed along with its limitations and plans for future work and extensions. Summary and concluding remarks are mentioned in Section \ref{sec:conclusion}.





\section{Background and Preliminaries}\label{sec:background}

\textbf{Integration testing:} testing can be done at different levels of abstraction and with respect to different software artifacts. Integration testing (considering both software and hardware) is defined by IEEE as `testing in which software components, hardware components, or both are combined and tested to evaluate the interaction between them' \cite{IEEE2017}. As some examples of catastrophic integration issues the following incidents can be named: i) explosion of Ariane 5 rocket in 1996 only seconds after its launch. The failure was caused due to data conversion from 64-bit floating point to 16-bit signed integer value, while the original value was greater than what a 16-bit signed integer could represent \cite{Ariane5}. The failure was attributed to lack of sufficient testing of reused software components \cite{Antonia}; ii) loss of NASA Mars orbiter costing \$125 million due to mismatch in the use of measurement units (conventional metric vs English) by different teams developing different parts of the system which when integrated and supposed to work together failed \cite{MarsOrbiter}.

There are already de facto tools, well-known frameworks, and automated solutions for unit testing such as JUnit, xUnit.net, and NUnit. However, this is not the case when it comes to integration testing \cite{CouplingPSO}. Also when we look into the research done in the literature on integration testing techniques, majority of them are either manual techniques, do not come with any tool support, or do not scale well to be applicable in real-world and industrial contexts. Moreover, many of them do not address the peculiarities of object-oriented applications \cite{AlexanderOffutt, Labiche:2000}.

Testing the integration of two or more components is generally more difficult than testing a single component (as in unit testing). There are certain challenges in integration testing, some of which are briefly touched upon here: 
\begin{itemize}
\item integration testing is usually done late in the development process,
\item awareness of emergent properties meaning that the properties of the whole are not necessarily the set or sum of the properties of the parts and components,
\item it is not trivial to simulate the environment and context of an integration prior to the actual integration, 
\item there can be many different combinatorial integration and interaction scenarios which need to be identified,
\item development of stubs and drivers when not all parts and components are in place,
\item challenge of commercial off-the-shell (COTS), or outsourced components; considering that different components may have been developed by different people with different assumptions.
\end{itemize}


\textbf{Coupling-based integration testing:} the main purpose in integration testing is to check whether different software components interact correctly with each other or not. Coupling-based testing is an integration testing technique in which different ways that two units interact with respect to their shared data and data flow are identified, and considered as the test requirements for test cases to fulfill \cite{Delamaro,Coupling}. In essence, the goal is to ensure that, with respect to the interface between two units, the variables \textit{defined} (i.e., assignment and storing a value for a variable in memory) in caller units are appropriately \textit{used} (i.e., access and reading of a variable in a statement) in the callee units. As described in \cite{Coupling}, coupling between two units measures the dependency relations between them by reflecting the interconnections that exist between the units. In a coupling relationship, faults in one unit may affect the coupled unit. Therefore, the stronger the coupling between the units, the higher can be the likelihood that a fault in one unit affects the other ones. 

In \cite{Offutt12}, Offutt et al. initially described 12 distinct and ordered coupling levels between two units/modules where for each coupling level the parameters are classified by the way they are used. These 12 levels were later grouped and combined into four (unordered) coupling types for testing purposes \cite{Coupling}: Call coupling, Parameter Coupling, Shared-data Coupling, External device coupling. Parameter coupling refers to passing of parameters between units. In the context of object-oriented applications, parameter coupling occurs when a method of one class uses and relies on the object of another class in its arguments \cite{WaheedQamar}. Considering the type of coupling between two units, and where a variable is defined in one unit and where it is used in the other one, a coupling path can be formed. These coupling paths are then considered as test requirements to be \emph{covered}. The main challenge in application of this technique is how to automate identification of coupling relationships, performing data flow analysis and pinpointing where a variable is defined and where it is used. Performing such tasks manually is not only cumbersome (if not impossible in large code bases) but also prone to human error, especially when they need to be repeated over and over again due to changes or committing new code. 


\textbf{Roslyn:} .NET compiler platform codenamed as project Roslyn provides an open-source set of compilers and code analysis APIs for C\# and Visual Basic. By opening up the capabilities of the compiler to the developers through this set of APIs, Roslyn establishes the concept of \textit{compilers as platforms}. In other words, Roslyn exposes the code analysis features of the compiler to a consumer through its API layer \cite{Roslyn1}. This enables to perform sophisticated code analysis by understanding the syntax, structure as well as semantics of the code. In general, Roslyn facilitates and provides for features such as meta-programming, code analysis, code synthesis, validation and enforcing of coding standards, making code fixes, etc.


\textbf{System Installer (SI):} 800xA is a distributed control system developed by ABB and a part of the ABB Ability\texttrademark~platform which is widely used in process control industry \cite{800xA}; such as pulp and paper, oil and gas industries, mining, chemical factories, and so on. System Installer is an application on 800xA which is responsible for delivery of new software, and updating and upgrading of existing components. System Installer which is implemented in C\# consists of many classes that constitute different functionality of the application in terms of network connectivity, file transfer, security and checking user permissions, software components installation, error management and exception handling, etc. System components are developed and undergo unit testing by different development teams and are committed to a central repository. Testing of these components at higher levels such as integration can be done by teams located in different parts of the world. Generally in such environments, when a problem at integration (or higher) level is detected, it should be forwarded back to the developers while they might have already moved on to developing other features, components, and parts of the system. This can incur huge cost and effort in finding and resolving integration bugs. Such situations can be alleviated if different interaction scenarios between components can be identified and exercised automatically, for instance as part of the nightly builds and testing process and before involving remote higher-level testing teams.


\section{Proposed Approach in IntegrationDistiller}\label{sec:approach}


	


In this section we describe the details of the IntegrationDistiller tool and the algorithms implemented as part of it; whose plan we originally presented in \cite{QRSRoslyn}. The algorithm described below is based on the concept of using method call sequences as integration test cases which is presented in \cite{WaheedQamar}. Basically the idea is to target coupling methods to exercise the integration and dependency of different classes while applying definition-use analysis to consider different method call sequences resulting in different states of class objects. The effectiveness of integration test cases generated using definition-use data flow analysis to capture different integration scenarios is already evaluated and demonstrated using mutation testing technique in \cite{IntegrationMutation}; and in \cite{Coupling} also a preliminary evaluation on the effectiveness of coupling-based testing is presented.

\subsection{Automatic generation of integration test paths}
One of the main features of IntegrationDistiller is the ability to automatically generate integration test paths by analyzing the code and identifying coupling methods. Coupling methods are class methods which have objects of other classes as part of their input parameters (parameter coupling). To construct coupling paths it is necessary to perform data flow analysis and identify definitions and uses of variables in callers and call sites respectively. This is basically done since based on the values assigned to class fields, the state of a class object can be different. Therefore, the data flow analysis is needed to identify sequence of method calls that can lead to different states of a class object in terms of the values of class fields used in the coupling method. In other words, while the parameter coupling captures the interaction and dependencies between classes, the order and sequence of method calls of a class can also result in additional combinatorial interaction scenarios to be tested. Considering this, the test path generation algorithm of IntegrationDistiller includes the following steps, inspired from the approaches described in \cite{WaheedQamar, IntegrationMutation}:

\begin{enumerate}
\item Read in as input the source code of the application and identify all (user-defined) classes
\item Identify coupling methods and their method signature from the methods of the classes identified in Step 1
\item Identify list of class field variables (attributes) that are used in the body of each coupling method of a class
\item Identify other class methods that can modify the value of the previously identified class fields
\item Identify list of class fields that are used in the body of these methods as well
\item Perform the last 2 steps recursively until all the methods in that class that modify the collected class fields and variables are identified
\item For each coupling method, construct a path whose end node is the coupling method, and the list of methods that can affect its field variables recursively as previous/children nodes of the path leading to that coupling method
\item As part of the path, add as nodes the instantiation of class objects where a new class object is accessed and needed. 

\end{enumerate}


Result of the above steps is a set of test paths in the form of method calls that cover coupling methods, and therefore, exercise the interaction among different classes based on parameter coupling. To enable generation of test paths in an automatic fashion the above steps need to be executed automatically. This is, however, not a trivial task and is a huge challenge in itself, as it requires careful analysis of the code, taking into consideration the abstract syntax tree, and semantics of code statements and blocks. Particularly, it is needed to automatically detect: class definitions, class methods, class method signatures and parameters, class fields, and also class fields definitions and uses in the body of each class method. These features and performing the aforementioned data flow analyses are automated as part of the code analysis engine of IntegrationDistiller based on Roslyn features and APIs which provide access to the capabilities of the compiler. The whole test generation process is incorporated and embedded with the development process so that new paths and thus test cases are generated automatically for new classes and code. So when new code is committed to the code repository or some modifications are made, IntegrationDistiller can be triggered automatically to generate new integration test paths covering newly added/modified code.

\subsection{Integration analytics and invocation points}
Besides the generation of integration test cases, IntegrationDistiller also provides insight into the code in terms of analytics on couplings and structures of classes and methods. This information is important and helpful since the way that application code and design are structured into a set of collaborating components and classes will have impact on complexity of integration testing of the application. Therefore, while the code analysis engine of IntegrationDistiller analyzes the code for test paths generation, the following integration metrics are also collected: 

\begin{itemize}
	\item Coupling degree of a class: total number of other classes that a class is dependent on. This is calculated as the total number of distinct types of class objects used in input parameters of a class methods. 
	\item Most used class in the application: shows which class is used the most by other classes as parameter coupling; i.e., by counting how many times a class type is used as a parameter type in the methods of all the classes. 
	\item Number of a class basetypes: showing if and how many other classes, a class inherits from.
	\item Length of test paths: for each test path generated, its length in terms of methods constituting nodes of the path are also calculated. 
	\item Other metrics: total number of methods of a class, number of constructors in a class.
\end{itemize}

Another concept that is considered in the code analysis engine of IntegrationDistiller is \textit{invocation points}. An invocation point is simply defined in our approach as the code statement in which a call to a method of another class is made (call site). Invocation points are also automatically detected in our approach, and total number of invocation points in each class method is calculated as another metric indicating reliance and dependency of a class and its methods on other classes. Moreover, it is also shown how many of the invocation points in a method are calls to user-defined classes, and how many are to native .NET classes (such as \texttt{Console}). It should be noted that the calculation and collection of all these analytics and metrics are also automated in IntegrationDistiller using Roslyn APIs, which can be repeated and performed easily again whenever a change is made in the code. 

\subsection{Code instrumentation and testing for timing properties}
Automatic detection of invocation points also enables IntegrationDistiller to automatically instrument the code at the right places for the purpose of testing the application for timing issues at integration level. If the code instrumentation is set to 'enabled' in IntegrationDistiller, when an invocation point is detected, code statements to log timestamps right before and after the identified invocation points are automatically added. This way, the time it takes to make a call to a method of another class is logged, and then it can be determined if such time durations are acceptable and desired, or violate some timing constraints. It is important to note that while there exist performance analytics and profiling tools (e.g., similar to gprof \cite{Graham:1982}) that may provide similar insights in terms of timing, the solution provided here is more on providing the general ability to automatically instrument the code at the right places for analysis of integration issues, which is customized in this work for timing measurements at integration level.

\section{Demonstration Example}\label{sec:example}
In this section, through an example we show how the proposed approach implemented in IntegrationDistiller works and what outputs are produced by it. Since we cannot include the actual source code of the ABB System Installer as our industrial use-case in the paper, and to have an easier code to follow, we use a more compact yet relatively complex example in terms of data flow with several levels of dependency to show the applicability of the approach. Figure \ref{lst:dummycode} illustrates a dummy code in which three classes are defined. As can be seen in the code, there is a coupling relationship between \texttt{Class A} and \texttt{Class B}, since methods \texttt{BM1} and \texttt{BM2} of \texttt{Class B} require an object of \texttt{Class A} in their parameters. Similarly, \texttt{Class C} has coupling to both \texttt{Class A} and \texttt{B}.

\begin{figure}[!htb]
{\scriptsize
\begin{Verbatim}[frame=single, numbers=left,numbersep=2pt]
class A
{public int x;}

class B:A
{
	public B(){ x = 7; }
	public B(int i){x = x + i;}
	public int Add(int j)
	{
		x= x + j;
		Console.WriteLine("{0}",x);
		return x;
	}
	void BM1(int test,A a)
	{
			Console.Write("dummy");
	}
	
	void BM2(int test,A a, int x1, int x2, int x3)
	{ ...
	}
}

class C
{
	private int var1, var2, var3;
	private int var4, var5;
	public C(B b)  // Coupling in constructor
	{
		var1 = b.x;
	}
	void CM1()
	{
		var5 = var4 + 2;
	}		
	void CM2()
	{
		var4 = var5 + 1;
		Console.WriteLine();
	}
	void CM3()
	{
		var4 = 10;
	}
	void  CM4()
	{
		var4 = 12;
		Console.WriteLine();
	}
	void CM5()
	{
		var3 = 127;
	}
	public int CM6(int k)
	{
		int ran1;
		B b1 = new B();
		ran1= b1.Add(2);  
		ran1 = b1.Add(ran1); 
		this.CM5();   
		C c3 = new C(b1);
		c3.CM4();     
		var1 = 5 * var4;		
		var2 = 3 * var3;
		return var1;
	}
	public void CM7(B b, A a)
	{
		int incr=0;
		++incr;
		int i1;
		i1 = var1 + 5;
		int i2;
		i2 = var2 + var3;
	}
}
\end{Verbatim}
}
\vspace{-.3cm}
\caption{Dummy code example}
\label{lst:dummycode}
\vspace{-.3cm}
\end{figure}

\begin{figure}[!htb]
{\scriptsize
\begin{Verbatim}[frame=single, numbers=left,numbersep=2pt]
Test Path Number: 1 ----- Path Length:1
	B:BM1(int test,A a)
Test Path Number: 2 ----- Path Length:1
	B:BM2(int test,A a, int x1, int x2, int x3)
Test Path Number: 3 ----- Path Length:3
	C:CM7(B b, A a)
	C:CM6(int k)
	C:CM5()
Test Path Number: 4 ----- Path Length:5
	C:CM7(B b, A a)
	C:CM6(int k)
	C:CM2()
	C:CM1()
	C:CM3()
Test Path Number: 5 ----- Path Length:5
	C:CM7(B b, A a)
	C:CM6(int k)
	C:CM2()
	C:CM1()
	C:CM4()
Test Path Number: 6 ----- Path Length:3
	C:CM7(B b, A a)
	C:CM6(int k)
	C:CM3()
Test Path Number: 7 ----- Path Length:3
	C:CM7(B b, A a)
	C:CM6(int k)
	C:CM4()
Test Path Number: 8 ----- Path Length:2
	C:CM7(B b, A a)
	C:CM5()
************* Constructors **********
Test Path Number: 9
	C:      C(B b)  
\end{Verbatim}
}
\vspace{-.3cm}
\caption{Generated test paths}
\label{lst:testpaths}
\vspace{-.3cm}
\end{figure}

\begin{figure}[!htb]
\centering
\includegraphics[width=\columnwidth]{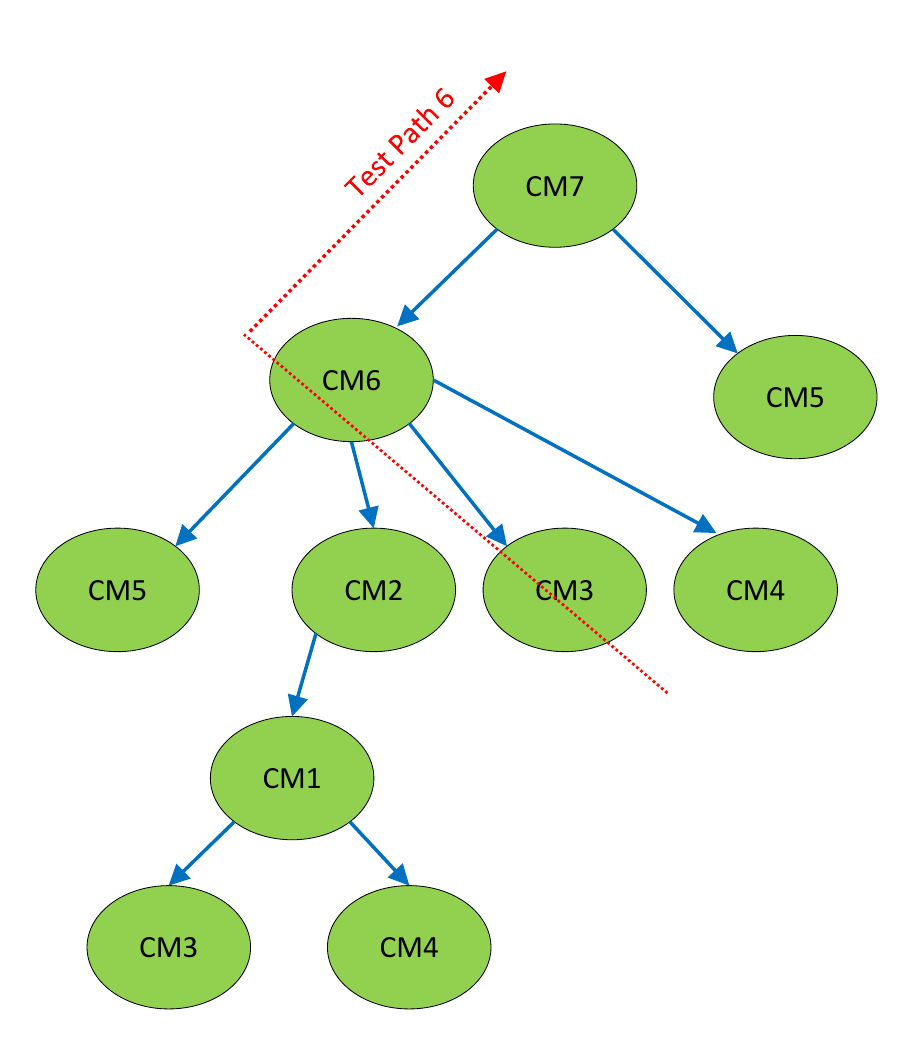}
\caption{Tree representation of test paths}
\label{fig:tree}
\end{figure}

By applying IntegrationDistiller to generate test paths for this code example, we get the following results as shown in Figure \ref{lst:testpaths}. Test Path 1 and 2 in Figure \ref{lst:testpaths} do not need much explanation as they simply exercise the dependency of \texttt{Class B} on \texttt{Class A} with respect to methods \texttt{BM1} and \texttt{BM2} respectively. In analysis of \texttt{Class C}, method \texttt{CM7} is identified as a coupling method. When an analysis of \textit{used} class fields (i.e., \texttt{var1}, \texttt{var2}, \texttt{var3} at lines 72 and 74 from Figure \ref{lst:dummycode}) is done on this method, it is identified that there are other methods in \texttt{Class C} that \textit{define} the values of those used class fields, namely methods \texttt{CM6} (at lines 63 and 64) and \texttt{CM5} (at line 52). These latter methods are considered as child nodes for the coupling method \texttt{CM7} in a tree structure. The same process is then repeated recursively for these methods as well: e.g., method \texttt{CM6} is analyzed to identify class fields which are used in its body, then other methods of \texttt{Class C} which modify those class fields are identified (\texttt{CM5}, \texttt{CM3}, \texttt{CM2}). Therefore in path 3 in Figure \ref{lst:testpaths}, we see that \texttt{CM6} is added as a child node for \texttt{CM7} and then since in the body of \texttt{CM6}, \texttt{var3} is used (line 64 in Figure \ref{lst:dummycode}) which is defined and modified in \texttt{CM5} (line 52), hence \texttt{CM5} is added as a child node for \texttt{CM6}. Finally, since there are no class fields that are used in the body of \texttt{CM5}, no further node is added to this path, the path is finalized, and the algorithm moves on to constructing test paths based on the other methods that define the variables used in \texttt{CM6} (i.e., \texttt{CM3}, \texttt{CM2}). Path 4 in Figure \ref{lst:testpaths} shows that in the analysis of method \texttt{CM2}, some used class fields are identified (\texttt{var5}) which are defined in \texttt{CM1}, which in turn includes used class fields (\texttt{var4}) which are defined in \texttt{CM3}. As a result, a tree representing test paths is constructed, part of which is shown in Figure \ref{fig:tree}. Execution of each test path is then done from the leaf nodes to the root node. For instance, in case of Test Path 6 (Figure \ref{lst:testpaths}), the order for execution will be: CM3{$\rightarrow$}CM6{$\rightarrow$}CM7.

Additionally class constructors are also analyzed for coupling parameters as can be seen at lines 32-34 in Figure \ref{lst:testpaths}. The analysis engine in IntegrationDistiller also produces detailed log information on the definition-use analysis of class fields as shown in Figure \ref{lst:analysislog}.

\begin{figure}[!htb]
{\tiny
\begin{Verbatim}[frame=single, numbers=left,numbersep=2pt]
From CM7 due to used variable:var1 --> CM6 which defines this variable.
From CM7 due to used variable:var3 --> CM5 which defines this variable.
From CM6 due to used variable:var3 --> CM5 which defines this variable.
From CM6 due to used variable:var4 --> CM2 which defines this variable.
From CM6 due to used variable:var4 --> CM3 which defines this variable.
From CM6 due to used variable:var4 --> CM4 which defines this variable.
From CM2 due to used variable:var5 --> CM1 which defines this variable.
From CM1 due to used variable:var4 --> CM3 which defines this variable.
From CM1 due to used variable:var4 --> CM4 which defines this variable. 
\end{Verbatim}
}
\vspace{-.3cm}
\caption{Definition-Use analysis log of class fields}
\label{lst:analysislog}
\vspace{-.3cm}
\end{figure}

As mentioned before, IntigrationDistiller is also capable of detecting invocation points and distinguishing which ones are calls to user-defined classes and which ones to native .NET classes. Part of the information produced on invocation points is shown in Figure \ref{lst:IntegrationPoints}. It shows that an invocation point is detected at line 48 (from Figure \ref{lst:dummycode}), which is a call to the \texttt{WriteLine} method of the \texttt{Console} class, hence it is marked that this is not a user-defined class. On the other hand, at line 58 another invocation point is detected due to a call to the \texttt{Add} method of \texttt{Class B}. In addition, the total number of invocation points detected in a class as well as its breakdown to the number of invocations per class methods are provided as well (lines 18-25 in Figure \ref{lst:IntegrationPoints}). 

\begin{figure}[!htb]
{\tiny
\begin{Verbatim}[frame=single, numbers=left,numbersep=2pt]
---- Invocations---
       *Invocation Point Detected at Line:48*
       Console.WriteLine()
       Invocation Class:Console      Not a user-defined class!        
       Current Class:C - In method:CM4
       Class object instance on which invocation detected:Console
				
       *Invocation Point Detected at Line:58*
       b1.Add(2)
       Invocation Class:B
       Current Class:C - In method:CM6
       Class object instance on which invocation detected:b1
...
---- End of Invocation analysis----
.
.
.
Number of invocation points in class C: 4 -- out of which 2 are User-Defined
  Number of invocation points in method CM1 is 0; out of which 0 are User-Defined
  Number of invocation points in method CM2 is 1; out of which 0 are User-Defined
  Number of invocation points in method CM3 is 0; out of which 0 are User-Defined
  Number of invocation points in method CM4 is 1; out of which 0 are User-Defined
  Number of invocation points in method CM5 is 0; out of which 0 are User-Defined
  Number of invocation points in method CM6 is 2; out of which 2 are User-Defined
  Number of invocation points in method CM7 is 0; out of which 0 are User-Defined
\end{Verbatim}
}
\vspace{-.3cm}
\caption{Invocation points analysis}
\label{lst:IntegrationPoints}
\vspace{-.1cm}
\end{figure}

By identifying the invocation points, IntegrationDistiller can \textit{automatically} add custom instrumentation code at the invocation points. This feature can be used to check for and log timing properties. Figure \ref{lst:instrumentation} shows an example instrumentation code that simply keeps the timestamps before and after the invocation at the invocation point (i.e., line 2: \texttt{ran1 = b1.Add(2);}) and writes out the time difference. It can then be decided if this time value is acceptable or not, and further analysis can be done to investigate the root cause of such timing violations across classes. In other words, while it does not per se show what has contributed and led to a timing violation, it helps to identify violation of timing properties when a method call needs to be done and return within a certain time period. In this regard, it also helps to get some idea about the vicinity of a timing violation in the code by knowing where to start investigating for the cause of the timing violation, rather than only noticing and observing a timing violation later and at a higher level for the whole application. 

\begin{figure}[!htb]
{\scriptsize
\begin{Verbatim}[frame=single, numbers=left,numbersep=2pt]
DateTime start_time3 = DateTime.Now;
ran1 = b1.Add(2);
TimeSpan timeDiff3 = DateTime.Now - start_time3;
Console.WriteLine("Line {0} took {1}"
		,58, timeDiff3.TotalMilliseconds);
\end{Verbatim}
}
\vspace{-.3cm}
\caption{Example code instrumentation to check for timing}
\label{lst:instrumentation}
\vspace{-.1cm}
\end{figure}

Some of the analytics results and metrics that are collected during the analysis of the code are shown in Figure \ref{lst:analytics}.


\begin{figure}[!htb]
{\scriptsize
\begin{Verbatim}[frame=single, numbers=left,numbersep=2pt]
-----
Class B
  Number of methods in Class B: 3
  Number of constructors in Class B: 2
  Maximum number of parameters among methods of class B: 5
  Coupling Degree of Class B: 1
  Bases of class B: A
  Number of base types of class B: 1
...
-----
Most used class: A
  3 times as method parameter
  0 times as variable type inside methods
\end{Verbatim}
}
\vspace{-.3cm}
\caption{Code metrics and analytics}
\label{lst:analytics}
\vspace{-.3cm}
\end{figure}

To give an idea about the performance of IntegrationDistiller, we have done an evaluation of its performance with respect to the above example on an Intel i7-5600U @2.60GHz machine with 12GB of RAM and Windows 10 Pro as the operating system. The results are summarized in Table \ref{tab:performance} (the values are in milliseconds, measured and converted using \texttt{System.DateTime} in .NET).

\begin{table}[htbp]
\begin{center}
\begin{scriptsize}
\begin{tabular}{|c|c|c|c|c|c}
\hline
\textbf{Feature} & \textbf{Run 1} & \textbf{Run 2} & \textbf{Run 3} & \textbf{Average} \\
\hline
Finding coupling methods &  3.9982  & 3.9977 & 4.0001 & 3.99867 \\
\hline
Integration \& coupling analytics & 5.0549 &	5.2171 & 4.5129 & 4.9283 \\
\hline
Test case generation & 5.9763	& 7.0152 & 6.9998 & 6.66377  \\
\hline
Invocation analysis & 781.5538 & 775.555 & 775.5745 &	777.5611 \\
\hline
Code Instrumentation & 148.9147	& 149.9162	& 148.8965	& 149.24247 \\
\hline
Invocation Analysis Per Class	& 2.997 & 1.9983 & 2.9997 & 2.665 \\
\hline
\end{tabular}
\end{scriptsize}
\end{center}
\caption{Performance evaluation on the example code}
\label{tab:performance}
\end{table}


As can be seen from the results, test case generation has taken about 6.66377 milliseconds on average for the above example. On the other hand, the invocation points analysis has been performed in about 777.5611 milliseconds on average. One reason that this activity and analysis has taken much more time than the others is that while for the other activities we only used the \textit{Syntax API} of Roslyn (for structural analysis of a program), for invocation points analysis we also needed make use of the \textit{Semantic API} as well to get and analyze the semantic model of the code, which is a more time-consuming type of analysis than structural. 

\section{Discussion and future work}\label{sec:discussion}

Identifying different ways that system components can interact is a huge challenge particularly when it is performed manually and with no tool support. Even if it may be feasible to do it manually for a small system or just one version of a large system, as soon as new changes are introduced, the same process needs to be performed again. Considering the size and complexity of industrial software applications, frequent changes and modifications (e.g., in agile and continuous development environments), as well as various versions and instantiations of a product that are developed, manual techniques for verification and validation of component interactions are not scalable and feasible anymore. IntegrationDistiller targets this problem by providing an automatic approach for identifying different interaction scenarios between class methods, and generating test paths as sequences of method calls. 

Use of Roslyn and .NET platform for implementation and automation of our approach have also enabled a more seamless integration of the testing process as part of the overall development process and with lower learning curve, since the System Installer application itself is also developed in .NET. On the other hand, while integration of a new testing process and framework into an already-established development process might in most cases be a one-time effort, however, issues such as maintenance, modifications, ease of use, configuration and setup, as well as impacts and changes in the daily routines and way-of-working of the developers should also be taken into account. As discussed, these issues can well hinder and prevent the adoption of (automated) testing techniques in industry \cite{Wiklund3480}.

One interesting feedback that we got from our industrial partner during the development of IntegrationDistiller was that the development team managers stated that the analytics information and metrics provided by the tool are not only helpful for testing purposes, but also are beneficial to the developers during the design and implementation phase. Developers can get better insight and find out, for instance: which components have highest degree of coupling and dependency, which methods are used and called the most, which are the most used classes, longest interaction paths and sequences, if class fields are defined/used correctly or in places where they are not supposed to, etc. This information can be useful in understanding the structure and design of the application and code in terms of integration scenarios and decide whether and where re-factoring might be needed and pay off the most. In addition, the most critical components of the system can also be identified whose failures may jeopardize the functionality of the whole application or many other components. Knowing this, it can be decided whether to allocate more testing efforts in order to more rigorously verify the correctness of such components.

The custom code instrumentation feature that we have implemented in IntegrationDistiller enables testers to automatically add custom code around invocation points for testing purposes. In this work, we used this feature mainly to get some time estimates for calls to methods of other classes (method invocations). Other factors such as the scheduling mechanism of the underlying platform and task preemptions \cite{monitoringCapabilities, accurateMonitoring} as well as the side effects of the instrumentation code itself should also be taken into account when more accurate estimates of timing properties are desired. The instrumented code, of course, needs to be executed in order to get the timing estimates and log them. On the other hand, all the instrumented code can easily be commented out or removed automatically, for instance after the debugging phase, if they are not to be part of the final application. Aside from timing properties, as a future work, it would be interesting to investigate if automatic code instrumentation can also be extended and used for testing of other extra-functional properties (EFPs)\footnote{Also referred to as non-functional properties (NFPs)} such as memory usage, energy consumption and the like as well. Finally, the code instrumentation feature can as well be exploited to automatically add exception handling code for method invocations.

Currently, our approach in IntegrationDistiller only considers parameter coupling. As another extension, including other types of coupling is also planned. Automatic creation of test scripts from test paths is another future extension of this work in order to increase the level of automation and cover test execution as well. It should be noted, however, that automating the testing process itself is in general a costly process and investment. Therefore, a careful cost-benefit analysis and evaluating where and when test automation can pay off, particularly in industrial contexts, is crucial \cite{GarousiAutomationMLR, Tahvili4438}. For instance, if a test is only executed once then perhaps investing to build an automatic test framework may not be worth the effort and cost.

The use of Roslyn for integration analysis and integration test case generation that we introduced in this paper is new. We intend to further extend the code analysis engine of IntegrationDistiller utilizing Roslyn features to provide more insight and metrics on the quality of applications, particularly with respect to integration scenarios. In this work, we addressed the scalability aspect through enabling effective automation. A more thorough scalability evaluation of the solution is planned as a future work.

\section{Summary and Conclusion}\label{sec:conclusion}
In this paper we discussed the importance and challenges of integration testing, as well as the characteristics that integration testing techniques should have in order to be usable and easier-to-adopt in industrial contexts. We introduced an approach for automatic integration analysis and generation of integration test cases for object-oriented systems, which is implemented in the IntegrationDistiller tool to target .NET applications in particular. The code analysis engine and test case generation approach in IntegrationDistiller is automated thanks to the features provided by the Roslyn compiler platform. Using Roslyn APIs we built a sophisticated code analysis engine that enables automatic generation of integration test paths, identification of invocation points in the code, and also automatic code instrumentation for further evaluation of interaction scenarios with respect to certain quality characteristics such as timing. This work has been initiated and motivated as the result of our collaboration with ABB Västerås and the observed challenges in integration testing of the 800xA System Installer application. While by no means this work alone answers all the integration testing challenges of our industrial partner in particular and industry in general, yet it serves as an example and step towards this goal, and also to facilitate knowledge transfer by making research solutions usable and applicable in industry. For instance, the definition-use data flow analysis that is automated in the analysis engine of IntegrationDistiller is a well-known technique in the literature, yet not prevalent and commonly used in industry (partly due to the lack of tool support and automation). However, IntegrationDistiller, among other things, enables testers to benefit from the result of this technique and apply it implicitly under the hood of its code analysis engine without the need to know much details about the technique itself.

\begin{spacing}{1}

\section*{Acknowledgements}\label{sec:am_acknowledgement}

This work has been supported by and received funding partially from the XIVT, TESTOMAT, and MegaM@Rt2 projects. We would like to also thank Henrik Strömblad and Jonas Stigeberg at ABB.

\bibliographystyle{IEEEtran}

\bibliography{all}

\end{spacing}

\end{document}